\documentclass[aps,prl,twocolumn,superscriptaddress,showpacs]{revtex4-1}
\usepackage{graphicx,color}
\definecolor{red}{rgb}{1,0,0}
\definecolor{green}{rgb}{0,1,0}

\begin{document}


\title{Nearly ferromagnetic metal state in the collapsed tetragonal phase of  YFe$_2$(Ge,Si)$_2$ }

\author{J. Srp\v{c}i\v{c}}
\affiliation{Jo\v{z}ef Stefan Institute, Jamova c. 39, 1000 Ljubljana, Slovenia}

\author{P. Jegli\v{c}}
\affiliation{Jo\v{z}ef Stefan Institute, Jamova c. 39, 1000 Ljubljana, Slovenia}

\author{I. Felner}
\affiliation{Racah Institute of Physics, The Hebrew University of Jerusalem, Jerusalem, 91904, Israel}

\author{Bing Lv}
\affiliation{Department of Physics, The University of Texas at Dallas, Richardson, Texas, 75080-3021, USA}

\author{C.W. Chu}
\affiliation{Department of Physics and the Texas Center for Superconductivity, University of Houston, Houston, Texas 77204-5005, USA}

\author{D. Ar\v{c}on}
\email{e-mail: denis.arcon@ijs.si}
\affiliation{Jo\v{z}ef Stefan Institute, Jamova c. 39, 1000 Ljubljana, Slovenia}
\affiliation{Faculty of Mathematics and Physics, University of Ljubljana, Jadranska c. 19, 1000 Ljubljana, Slovenia}


\begin{abstract}
The surprising discovery of tripling the superconducting critical temperature of KFe$_2$As$_2$ at high pressures issued an intriguing question of how the superconductivity in the collapsed tetragonal phase differs from that in the non-collapsed phases   of Fe-based superconductors. Here we report  $^{89}$Y nuclear magnetic resonance study of  YFe$_2$Ge$_{x}$Si$_{2-x}$ compounds whose electronic structure is similar to that of iron-pnictide collapsed tetragonal phases already at ambient pressure. 
Fe(Ge,Si) layers show  strong  ferromagnetic spin fluctuations whereas layers are coupled antiferromagnetically -- both positioning the studied family close to a quantum critical point. 
Next, localized moments attributed either to Fe interstitial  or antisite defects may account for magnetic impurity pair-breaking effects thus explaining the substantial variation of superconductivity among different YFe$_2$Ge$_2$ samples. 
\end{abstract}

\pacs{76.60.-k, 75.50.Cc, 73.43.Nq, 74.70.Xa}

\maketitle
%

The collapsed tetragonal phase (CTP) found in the family of $A$Fe$_2$As$_2$ ($A =$ Ba, Ca, Eu, Sr, K) at high pressures has been  considered as a non-superconducting phase \cite{PrattPRB}, because the formation of interlayer As-As  bonds triggers topological change of the Fermi surface thus removing for the superconductivity important nesting conditions \cite{ColdeaPRL}.   This notion has suddenly changed by the recent discovery of tripling the superconducting critical temperature $T_{\rm c}$ in KFe$_2$As$_2$ at pressures higher than $\sim 15$~GPa when CTP is formed \cite{NakajimaPRB, ChenHP}.  The strong electron correlations \cite{ChenHP} or almost perfectly nested electron and hole pockets found for KFe$_2$As$_2$ in CTP \cite{Valenti} were both put forward to explain the  surprising enhancement of $T_{\rm c}$.  Thus, to what degree the superconducting pairing mechanism of CTP differs from that of the non-collapsed layered Fe-based phases \cite{HosonoRev} remains at present unclear.

\begin{figure}[tbp]
\includegraphics[width=1.0\linewidth]{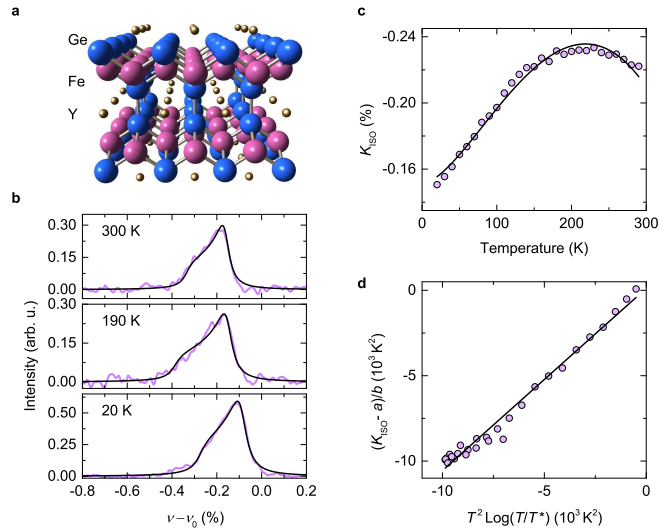}
\caption{\label{fig1} (color online). (a) The body-centered tetragonal crystal structure of  YFe$_2$Ge$_x$Si$_{2-x}$ (space group $I4/mmm$). Here small gray, large blue and  large pink spheres represent Y, Ge/Si and Fe atoms, respectively. (b) The $^{89}$Y NMR spectra (thick violet lines) of polycrystalline YFe$_2$Ge$_{0.2}$Si$_{1.8}$ samples at selected temperatures. Solid black line is a fit to uniaxial  shift anisotropy. (c) Temperature dependence of $^{89}$Y NMR shift, $K_{\rm iso}$ (circles). Solid  line is a fit to Eq. (\ref{chiLn}). (d) Semi-log plot of $K_{\rm iso}$ {\em vs.}  $T^2\ln (T/T^*)$  reveals a straight line thus corroborating with a nearly ferromagnetic metal state. }
\end{figure}

Rare earth iron silicides and germanides of the $R$Fe$_2$$X_2$ type ($R=$ rare earth element, $X=$ Ge, Si) have been studied since the 1970's for their magnetic properties -- various probes showed  no long-range magnetic order  in this family of materials \cite{Felner1975, NdFe2Si2}.   
The two representative compounds YFe$_2$Si$_2$ and YFe$_2$Ge$_2$ are  isostructural to $A$Fe$_2$As$_2$, i.e., they all grow in the same body-centered tetragonal crystal structure (Fig. \ref{fig1}a). 
The ratio of YFe$_2$Ge$_2$ tetragonal lattice parameters \cite{Ge2SC}  is $c/a=2.638$, which is very close to $c/a\approx 2.5$ of the high-pressure CTP in  KFe$_2$As$_2$ \cite{ChenHP}. The structural resemblance with CTP of  KFe$_2$As$_2$ is reflected  in the  similarities of the electronic structures of the two compounds \cite{Valenti, LDA, GGA, ChenPRL2016}. Because of the collapsed tetragonal structure, the  interlayer Ge-Ge bonds make the band structure and the Fermi surfaces of YFe$_2$Ge$_2$ more three-dimensional \cite{GGA}. While nesting of hole and electron pockets, similar to KFe$_2$As$_2$, may still imply that putative superconductivity in YFe$_2$Ge$_2$ has the standard $s_{\pm}$ order \cite{LDA}, the ferromagnetic spin fluctuations  within Fe-layers could promote even triplet superconductivity \cite{GGA}. Reports on experimental observation of superconductivity are equally controversial. 
Superconductivity has been initially  reported for YFe$_2$Ge$_2$ below $T_{\rm c}=1.8$~K \cite{Ge2SC}. 
On the other hand, no bulk superconductivity  down to 1.2~K was observed in \cite{ Felner2015}  and  it was argued that the superconductivity has a filamentary nature \cite{KimPhil2015}.  However, more recent studies claimed bulk superconductivity in high quality YFe$_2$Ge$_2$ ingots with $T_{\rm c}$ strongly dependent on the sample quality \cite{ChenPRL2016}.

The key to  understanding  such conflicting findings is hidden in the normal state of YFe$_2$Ge$_{x}$Si$_{2-x}$ family. First principle calculations \cite{LDA, GGA, Singh2016} for YFe$_2$Ge$_2$ and YFe$_2$Si$_2$ suggested that the favorable three-dimensional magnetic order is antiferromagnetic stacking of ferromagnetic Fe-layers along the tetragonal $c$-axis.  However, such long-range antiferromagnetic order has never been experimentally observed despite the  strongly enhanced spin susceptibility in the normal state \cite{Avila, Felner2014, Felner2015}.  We note though that the sister LuFe$_2$Ge$_2$ compound  complies with the proposed antiferromagnetic ordering at $T_{\rm N}=9$~K \cite{Avila}. Hence, it has been suggested that YFe$_2$Ge$_2$ is very close to an antiferromagnetic quantum critical point \cite{Ge2SC, Felner2015, LDA, GGA}. Moreover, strong quantum spin fluctuations \cite{Sirica, Ferstl} and the maximum observed in the magnetization measurements across the whole family of YFe$_2$Ge$_{x}$Si$_{2-x}$  have been attributed to a nearly ferromagnetic metal state \cite{Felner2014, Felner2015}.
Therefore, if superconductivity in YFe$_2$Ge$_2$ is indeed intrinsic,  it develops from a quantum critical state where strong spin fluctuations probably play an important role in tuning $T_{\rm c}$.

%
 
Nuclear magnetic resonance (NMR) has been pivotal in studies of 
iron-based superconductors \cite{Grafe, Imai2008, Kawasaki, NOFA, NakaiNJP, LiFeAs, NaFeAs, iFeSe, Ma2012, Wiecki} as well as in studies of systems close to the quantum critical point \cite{CFNMR, CFNMR1, CFNMR2, CFNMR3, CFNMR4, Moriya1, Moriya2,  Cr2B}. Here, we employ  $^{89}$Y NMR to probe the normal state of  YFe$_2$Ge$_x$Si$_{2-x}$ compounds. 
Data reveals that YFe$_2$Ge$_x$Si$_{2-x}$ are indeed very close to the quantum critical point -- the $c/a$ ratio acting as a control parameter to tune the magnetism. In tetragonal Fe-based structures with $c/a\sim 2.5$, such as YFe$_2$Ge$_x$Si$_{2-x}$, CTP of KFe$_2$As$_2$  or SrCo$_2$(Ge$_{1-x}$P$_x$)$_2$ \cite{CavaNP}, then even small perturbations, like Fe interstitial  or antisite defects discovered in this work, can have profound effect on the adopted  state.  





The $^{89}$Y (nuclear spin $I=1/2$) NMR spectrum of YFe$_2$Ge$_{0.2}$Si$_{1.8}$ \cite{suppl}, taken at $T=300$~K, shows a single line with a characteristic powder pattern of an axially symmetric shift anisotropy (Fig. \ref{fig1}b). 
Excellent fitting of the spectrum is achieved with the isotropic part of the  shift $K_{\rm iso}=(2K_\perp + K_{||})/3 = -0.222$\% and the shift anisotropy $\delta K=K_\perp-K_{||}=0.174$\% ($K_\perp$ and $K_{||}$ are the two principal values of the $^{89}$Y NMR shift tensor $\textbf{K}$). 
In general, the $^{89}$Y NMR shift has two main contributions: the temperature independent chemical shift and the hyperfine shift. The former is for $^{89}$Y known to be on the order of $200$~ppm or even less \cite{Ychem}, so we conclude that the main contribution to $\textbf{K}$ arises from the hyperfine interactions of $^{89}$Y with itinerant charges of Fe(Si,Ge) layer.  From $K_{\rm iso}={a_{\rm iso}\over N_{\rm A}\mu_{\rm B}}\chi$ ($N_{\rm A}$ and $\mu_{\rm B}$ are the Avogadro number and the Bohr magneton, respectively) and the room temperature value of spin susceptibility $\chi = 3.3\cdot 10^{-3}$~emu/mol we estimate  the isotropic hyperfine constant to be $a_{\rm iso}=-3.8$~kOe/$\mu_B$. This is by a factor of $\sim 2$ larger compared to, e.g.,   YBa$_2$Cu$_3$O$_{7-y}$ high-T$_c$ superconductors \cite{Taki, AlloulPRL1993} implying  strong coupling of yttrium layer to the itinerant charges in the electronically active Fe(Ge,Si) layer and consistent with a more three-dimensional band structure \cite{GGA}.
 
On cooling the $^{89}$Y NMR spectra retain their axially symmetric shift anisotropy lineshape at all temperatures (Fig. \ref{fig1}b). The  $^{89}$Y NMR line first slightly shifts to even more negative values of $K_{\rm iso}$ but then below $\sim$200~K the trend suddenly reverses and  the shift, and thus also the local spin susceptibility probed by $^{89}$Y, is significantly reduced compared to the room temperature value. The shift anisotropy follows the same trend, e.g., the most shifted spectrum at $T_{\rm max}=200$~K has  also the largest $\delta K$. The absence of any significant broadening of $^{89}$Y NMR spectra down to lowest temperatures  clearly rules out  long-range magnetic ordering in YFe$_2$Ge$_{0.2}$Si$_{1.8}$.

$K_{\rm iso}$  thus has a pronounced minimum (or maximum in $|K_{\rm iso}|$) at $T_{\rm max}$ (Fig. \ref{fig1}c) consistent with a maximum in the local spin susceptibility probed by  $^{89}$Y. Such non-monotonic dependence of $\chi$ strongly deviates from a simple Pauli paramagnetism in metals and is  suggestive of  spin correlations. The corrections to the temperature dependence of  spin susceptibility of normal paramagnetic metals in the presence of spin fluctuations close to the magnetic ordering have been a subject of intense theoretical discussions \cite{MonodPRL, Misawa1970,Barnea1975, Barnea1977}. The maximum in $\chi (T)$ is predicted as the spin susceptibility is given by 
\begin{equation}
\chi(T) = \chi(0) -bT^2\ln (T/T^*)\, ,
\label{chiLn}
\end{equation}
where $\chi(0)$ is the  Pauli spin susceptibility enhanced by enhancement factor $S$, $T^*$ reflects the cutoff  energies 
whereas prefactor $b$ is also strongly dependent on the enhancement factor, i.e., $b\propto S^4$. Inserting  Eq. (\ref{chiLn})  into the expression for $K_{\rm iso}$  yields $\chi(0) = 2.5(1)\cdot 10^{-3}$~emu/mol, $b=0.11(1)$~emu/(mol K$^2$) and $T^*=351(2)$~K. The agreement with the model is further demonstrated on a semi-log plot of $K_{\rm iso}$ {\em vs} $T^2\ln (T/T^*)$ where all experimental points fall on a straight line (Fig. \ref{fig1}d).  The electronic states around the Fermi level originate from the Fe $3d$ derived bands \cite{LDA, GGA, Singh2016} and within the first principle computations yield the bare density of states $N(0)\approx 4.5$~eV$^{-1}$. By comparing  $\chi(0)$ to the calculated Pauli susceptibility $\chi_{\rm P}=2.9\cdot 10^{-4}$~emu/mol we evaluate the large Stoner enhancement factor $S= 8.6$.    YFe$_2$Ge$_{0.2}$Si$_{1.8}$  thus  fits into a class of nearly  ferromagnetic metals \cite{Felner2014, Felner2015}.

\begin{figure}[tbp]
\includegraphics[width=1\linewidth]{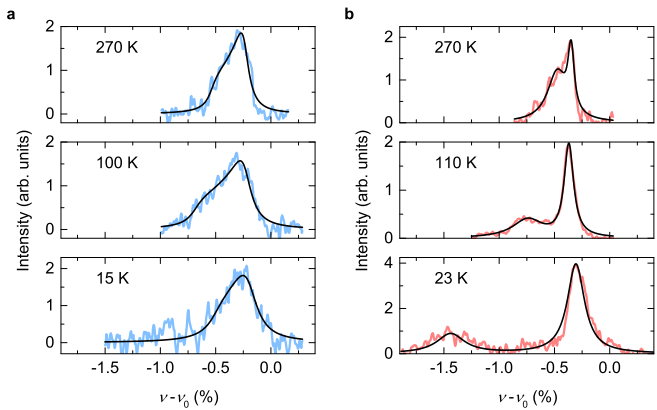}
\caption{\label{fig2} (color online).  $^{89}$Y NMR spectra of (a) YFe$_2$GeSi and (b) YFe$_2$Ge$_{2}$  shown for some selected temperatures. Thick lines represent the experimental data, while thinner solid lines are lineshape fits.  }
\end{figure}

Partial or complete replacement of Si with Ge yields isostructural  YFe$_2$GeSi and YFe$_2$Ge$_{2}$ compositions. 
Compared to YFe$_2$Ge$_{0.2}$Si$_{1.8}$, the $^{89}$Y NMR spectrum of YFe$_2$GeSi is significantly broader and shifted to even lower resonance frequencies (Fig. \ref{fig2}a). The broadening could be  attributed  to the effects of local site disorder introduced by a random  Si and Ge occupancy of $4e$ crystallographic positions. On the other hand, since it is  unlikely that the structural and electronic modifications within the Fe(Ge,Si) layer would considerably affect the values of $^{89}$Y hyperfine constant $a_{\rm iso}$, the observed monotonic increase of the $^{89}$Y shift  with increasing Ge content can only reflect the  enhancement of local spin susceptibilities by nearly a factor of $\sim 2$ when compared to  YFe$_2$Ge$_{0.2}$Si$_{1.8}$. The $^{89}$Y NMR spectrum of YFe$_2$Ge$_{2}$ (Fig. \ref{fig2}b) is shifted even more thus implying even larger local spin susceptibilities probed by $^{89}$Y NMR.

\begin{figure}[tbp]
\includegraphics[width=1\linewidth]{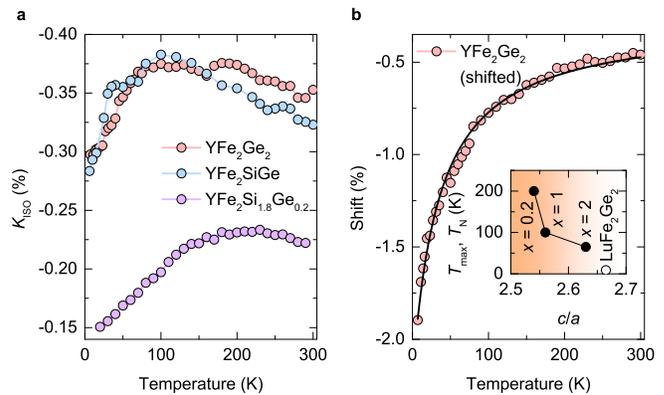}
\caption{\label{fig3} (color online). (a) Comparison of temperature dependences of the isotropic part of the $^{89}$Y NMR shifts $K_{\rm iso}$ measured for YFe$_2$Ge$_{0.2}$Si$_{1.8}$ (violet), YFe$_2$GeSi (blue)  and  YFe$_2$Ge$_{2}$ (red)  powders. (b) The shift of the additional $^{89}$Y resonance (circles), with the signal intensity of about 20\% of the total NMR signal, follows a Curie-like temperature dependence (solid line).  Inset: The dependence of temperature $T_{\rm max}$, i.e., temperature where $|K_{\rm iso}|$ has a maximum for YFe$_2$Ge$_x$Si$_{2-x}$ family, on the $c/a$ ratio (solid circles). The open circle stands for $T_{\rm N}=9$~K of LuFe$_2$Ge$_2$ \cite{Avila}.  }
\end{figure}

$^{89}$Y NMR spectra retain their axially symmetric shift anisotropy lineshape at all temperatures (Fig. \ref{fig2}) thus indicating that there is no structural phase transition between 300 and 15~K that would reduce the $^{89}$Y $2a$ site symmetry in both samples. Compared to  YFe$_2$Ge$_{0.2}$Si$_{1.8}$  the temperature $T_{\rm max}$ where $K_{\rm iso}$ reaches its minimum (maximum in $|K_{\rm iso}|$) is systematically reduced with increasing Ge content (Fig. \ref{fig3}a), i.e. to $\sim 100$~K and $\sim 70$~K in  YFe$_2$GeSi and  YFe$_2$Ge$_{2}$ samples (inset to Fig. \ref{fig3}b), respectively. Moreover, fitting of the temperature dependences of $K_{\rm iso}$ to Eq. \ref{chiLn} is not satisfactory anymore. Even extensions of paramagnon models for nearly ferromagnetic metals to take into account effects of impurities \cite{Barnea1977}, i.e.,  $\chi(T) = \chi(0) -bT^2\ln [(T+T_{\rm imp})/T^*]$ where $T_{\rm imp}$ is related to the effects of finite mean free path on the spin fluctuations, does not improve the quality of the fit. Contrary to YFe$_2$Ge$_{0.2}$Si$_{1.8}$, the $^{89}$Y NMR spectra remain broad even when $|K_{\rm iso}|$ is reduced at low temperatures (e.g., compare the spectra of YFe$_2$Ge$_{2}$ measured at 110 and 23~K on Fig. \ref{fig2}b). This is indicative of the growth of local magnetic fields at $^{89}$Y sites probably originating from the  short-range static magnetic correlations that  begin to develop in a high magnetic field of 9.34~T at low temperatures. We note that $|K_{\rm iso}|$ is suppressed at lowest temperatures, which necessitates that correlations between the nearly ferromagnetic Fe(Si,Ge) layers are of antiferromagnetic nature.   Although still no long-range order is established, the analysis proves a systematic evolution of magnetism in YFe$_2$Ge$_x$Si$_{2-x}$ where nearly ferromagnetic metal layers for $x=0$ becomes progressively more antiferromagnetically coupled as $x\rightarrow 2$.

\begin{figure}[tbp]
\includegraphics[width=1.0\linewidth]{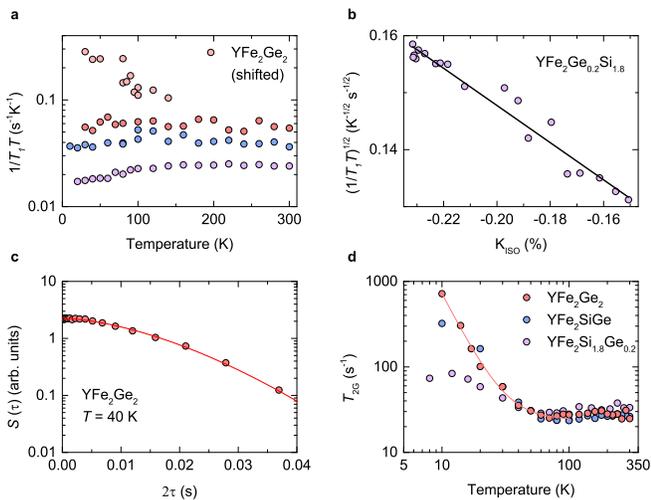}
\caption{\label{fig4} (color online). (a) Temperature dependences of $^{89}$Y spin-lattice relaxation rates $1/T_1T$ for YFe$_2$Ge$_x$Si$_{2-x}$ samples. (b) Test of the Korringa relation for YFe$_2$Ge$_{0.2}$Si$_{1.8}$  by plotting   $\sqrt{1/T_1T}$ {\em vs.} $K_{\rm iso}$ with temperature as an implicit parameter. 
(c) The decay of echo signal intensity as a function of interpulse delay time $\tau$ measured in YFe$_2$Ge$_2$ at $T=40$~K. The solid line is a fit with $\alpha = 1.65$ (see text for details).
(d) Temperature dependences of Gaussian spin-spin relaxation rates $1/T_{2\rm G}$. Solid red line is a fit for YFe$_2$Ge$_2$ to a low-temperature power-law $T^{-n}$ dependence with $n=2.9(1)$. The labeling of different samples is provided in insets.  
}
\end{figure}

Another peculiarity of  the YFe$_2$Ge$_{2}$ sample is a pronounced shoulder in the $^{89}$Y NMR spectra, which on cooling  develops into a separate resonance with an extremely large shift   (Fig. \ref{fig2}b). This  spectral component   is absent (or much weaker) in the other two compounds. 
Since all three studied samples grow in the same space group with a single crystallographic Y site there is no obvious reason for a separate $^{89}$Y NMR line in this case. The intensity of this signal is  about 20\% of the total $^{89}$Y NMR signal so  it  cannot be hastily attributed to some extrinsic  impurity phase leading us to the conclusion  that it must be intrinsic to YFe$_2$Ge$_{2}$. 
The temperature dependence of the shift (Fig. \ref{fig3}b) follows a perfect Curie-like dependence, i.e. $K_{\rm iso}^{\rm s}=K_0+C/(T-T_0)$ with $K_0 = -1890$~ppm, $C=-0.89$~K and $T_0=-48$~K thus associating this signal with $^{89}$Y sites located close to some localized moments. This notion is further supported by the measurements of the spin-lattice relaxation rate, $1/T_1$, which is for this component nearly temperature independent (Fig. \ref{fig4}a). 
To explain the presence of localized paramagnetic impurities we refer here to a common feature frequently encountered also in iron-pnictide and iron-chalcogenide samples \cite{iFeSeBlund, Klauss, iFeSe} , i.e., that some of Fe create antisite  (Fe occupying  Ge/Si site) or interstitial defects (Fe occupying  crystallographic interstitial sites between Y and Ge layers).
Due to the large moment of localized Fe defects, a strong hyperfine field with a Curie-like dependence is anticipated on the  nearest neighboring $^{89}$Y sites in agreement with the experiment.  
We stress that the presence of such localized moments can provide a very efficient channel for magnetic impurity pair-breaking effects thus explaining the large variation of  $T_{\rm c}$'s in  different YFe$_2$Ge$_{2}$ samples \cite{Ge2SC, Felner2015,  KimPhil2015, ChenPRL2016}.  

%

For systems close to an antiferromagnetic quantum critical point strong quantum spin fluctuations are responsible for a characteristic power-law dependence of spin-lattice relaxation rate, $1/T_1$, i.e., $1/T_1T\propto T^{-n}$ with $n=3/4$ \cite{Moriya1, Moriya2, CFNMR1, CFNMR3}. However, for YFe$_2$Ge$_{x}$Si$_{2-x}$ the respective  $^{89}$Y spin-lattice relaxation rates divided by temperature, $1/T_1T$, do not show such dependence (Fig. \ref{fig4}a). For YFe$_2$Ge$_{0.2}$Si$_{1.8}$, the temperature dependence of $1/T_1T$ resembles that of $K_{\rm iso}$, i.e., it exhibits a broad maximum at  $\sim 200$~K.
For correlated metals where electron-electron exchange enhancement effects are important, the Korringa relation reads \cite{Walstedt}
\begin{equation}
{T_1TK_{\rm iso}^2}= {\hbar \over 4\pi k_{\rm B}} {\gamma_{\rm e}^2\over \gamma_{89}^2}\beta\, .
\label{Korr}
\end{equation}
Here $\gamma_{\rm e}$ and $\gamma_{89}$ are the electronic and $^{89}$Y  gyromagnetic ratios, respectively. The  Korringa factor $\beta$ is introduced to account for the electron-electron exchange in a strongly correlated metal \cite{Walstedt, Stenger}. Plotting $\sqrt{1/T_1T}$ {\em vs.} $K_{\rm iso}$ we find the expected linear dependence (Fig. \ref{fig4}b) yielding $\beta =8.3$. Such strong enhancement is arising from the in-plane ferromagnetic fluctuations thus corroborating the nearly ferromagnetic metal state. The spin-lattice relaxation rates of YFe$_2$GeSi and YFe$_2$Ge$_{2}$ are even more enhanced implying even larger $\beta>15$. However, the nearly constant $1/T_1T$   does not scale with the temperature dependent $K_{\rm iso}^2$ giving a strong indication for a Fermi-liquid breakdown for these two samples. 

On the other hand, the effect of inter-plane antiferromagnetic spin fluctuations is clearly  absent in  $1/T_1$ data because such spin fluctuations are filtered out at the highly symmetric yttrium position. 
This explains why $1/T_1T$ is temperature independent for  YFe$_2$GeSi  and YFe$_2$Ge$_{2}$  despite the indications for antiferromagnetic correlations between layers given by the temperature dependence of $K_{\rm iso}$. 
To confirm antiferormagnetic correlations between layers we finally turn to  $^{89}$Y spin-spin relaxation rates, $1/T_2$. We first fitted  the envelope of the echo decay $S(\tau)$ as a function of the interpulse delay time $\tau$ to a phenomenological expression $S(\tau)=S_0 \exp [-(2\tau/T_2)^\alpha ]$ (Fig. \ref{fig4}c). Here $S_0$ is the initial echo intensity signal whereas the  parameter $\alpha$ expresses the relative contributions of the Redfield and the Gaussian part of the echo decay \cite{Walstedt, CurroT2G}. 
We obtain $\alpha\approx 1.6$, showing that both relaxation channels are present and comparable. 
Therefore,  in the next step we employed the procedure of \cite{CurroT2G} to extract the Gaussian part of the decay, $T_{2\rm G}$. As anticipated, $1/T_{2\rm G}$ is constant at high temperatures for all samples  (Fig. \ref{fig4}d). However, at low temperatures $1/T_{2\rm G}$ nearly diverges for YFe$_2$Ge$_2$ and YFe$_2$GeSi, whereas  it is only slightly enhanced for  YFe$_2$Ge$_{0.2}$Si$_{1.8}$. Below 100~K, $1/T_{2\rm G}$  is for YFe$_2$Ge$_2$ fitted to $1/T_{2\rm G}=1/T_2^0+BT^{-n}$ with  a power exponent $n=2.9(1)$. This is reminiscent to cuprates, where  Gaussian contribution to the echo decay is proportional to the antiferromagnetic correlation length $\xi$, i.e., $1/T_{2\rm G}\propto \xi$ \cite{CurroT2G, Walstedt, PinesT2G, PennT2G}.
The low-temperature enhancement in $1/T_{2\rm G}$ thus demonstrates the build-up of antiferromagnetic correlations between ferromagnetic Fe(Ge,Si) layers  proving the vicinity of antiferromagnetic quantum critical point in Ge-rich samples.

The YFe$_2$Ge$_x$Si$_{2-x}$ family displays  strong ferromagnetic  fluctuations within the Fe(Ge,Si) layers and antiferromagnetic correlations between layers. 
These correlations grow in importance with $x\rightarrow 2$ thus suggesting that with the introduction of slightly larger Ge ions the change in the $c/a$ ratio is sufficient ($c/a=2.54$, 2.56 and 2.64 for $x=0.2, 1\, {\rm and}\,  2$ samples, respectively)
 to strengthen the inter-layer coupling.  The trend is additionally supported by the long range  antiferromagnetic order below 9 K in LuFe$_2$Ge$_2$ ($c/a=2.66$) \cite{Avila}. 
YFe$_2$Ge$_2$ (and probably also  YFe$_2$GeSi) thus must be very close to the quantum critical point (inset to Fig. \ref{fig3}b) therefore accounting for the enhanced quantum spin fluctuations. 
We stress that a similar Ge-Ge bonding strength acting as a tuning parameter to induce the quantum critical point has been reported  for SrCo$_2$(Ge$_{2-x}$P$_x$)$_2$, which likewise belongs to the same layered tetragonal ThCr$_2$Si$_2$ structure type \cite{CavaNP}.
When quantum critical point separates the magnetic and superconducting phases, even small perturbations introduced by defect localized moments, such as those reported here, may have profound effect on the ground state.
Although there is no NMR data available for a comparison with the high-pressure CTP phase of KFe$_2$As$_2$, our results prove a fundamentally different normal state of YFe$_2$Ge$_x$Si$_{2-x}$  compared to that of the superconducting iron-pnictides. It is therefore unlikely that  superconductivity in YFe$_2$Ge$_2$ follows the same scenarios as those discussed for iron-pnictides \cite{HosonoRev}.




%
%
%
D.A.  acknowledges the financial support by the Slovenian Research Agency, grant No. N1-0052.  
B. L and C. W. C acknowledge the financial support by US Air Force Office of Scientific Research Grants No. FA9550-15-1-0236. 
\bibliography{YFGS}
\bibliographystyle{apsrev}

\end{document}